\documentclass[showpacs,preprintnumbers,amsmath,amssymb]{revtex4}
\usepackage{graphicx}
\usepackage{dcolumn}
\usepackage{bm}
\usepackage{float}
\usepackage{subfigure}
\usepackage{extarrows}
\usepackage{times}
\usepackage{threeparttable}

\begin{document}

\title{Influence of counter-rotating interaction on quantum phase transition in Dicke-Hubbard lattice: an extended coherent state approach}

\author{Yongchuan Lu$^{1}$}
\author{Chen Wang$^{1}$}\email{wangchenyifang@gmail.com}

\address{
$^{1}$Department of Physics, Hangzhou Dianzi University, Hangzhou, Zhejiang 310018, China}

\date{\today}

\begin{abstract}
We investigate the ground state behavior of the Dicke-Hubbard model including counter-rotating-terms.
By generalizing an extended coherent state approach within mean-field theory, we self-consistently obtain the ground state energy and delocalized order parameter.
Localization-delocalization quantum phase transition of photons is clearly observed by breaking the parity symmetry.
Particularly, Mott lobes are fully suppressed, and the delocalized order parameter shows monotonic enhancement by
increasing qubit-cavity coupling strength,
in sharp contrast to the Dicke-Hubbard model under rotating-wave approximation.
Moreover, the corresponding phase boundaries are stabilized by decreasing photon hopping strength, compared to the Rabi-Hubbard model.
\end{abstract}

\pacs{05.70.Fh, 03.65.Ud, 42.50.Pq}

\maketitle

\section{Introduction}
Deep understanding and smart control of quantum phase transition (QPT) of photons in light-matter interacting systems, one of classical phenomena in quantum world, is of fundamental importance~\cite{greentree2006quantum,hartmann2006strongly}.
QPT reveals quantum critical fluctuations at the ground state by tuning system parameters independent of temperature,
which was originally introduced in condensed matter physics to investigate strongly correlated effect of electrons~\cite{sachdev2011quantum}.
Recently, QPT has attracted intensive studies in realizable photon lattices, where photons are coherently excited and transferred~\cite{hartmann2008quantum}.
Light and matter equally contribute to unravel this novel effect.
Hence, This greatly broadens potential applications of photons to identify quantum criticality and coherence in many-body systems.

The simplest paradigm to describe photon lattices with light-matter interaction is known as Jaynes-Cummings-Hubbard (JCH) model,
which is composed by an array of single-mode photonic cavities, with each individually coupled to two-level system, e.g., qubit~\cite{greentree2006quantum,hartmann2006strongly,hartmann2008quantum,na2008strongly,koch2009superfluid,hur2015many,schmidt2009strong,schmidt2010excitations,
nietner2012ginzburg,you2014phase,bujnowski2014supersolid,hayward2015superfluid}.
The competition between the intra-site qubit-cavity coupling based on rotating-wave approximation (RWA) and inter-site photon hopping has been extensively investigated, which results in the nontrivial QPT.
Specifically, The mean-field theory has initially been introduced to discover the Mott-insulating-to-superfluid-like phase transition of light,
quantified by the excitation number of polaritons~\cite{greentree2006quantum,na2008strongly}.
The corresponding phase diagram shows analogy with that in the seminal Bose-Hubbard model~\cite{fisher1989boson,bloch2008many,lv2015coulomb}, where Mott lobes are clearly exhibited.
Consequently, numerical exact approaches have been employed to confirm phase boundaries~\cite{hartmann2008quantum,rossini2007mott} and provide critical behaviors~\cite{zhao2008insulator,pippan2009excitation,hohenadler2011dynamical}.
Moreover, the rapid progress of circuit-quantum electrodynamics and trapped ions have directly observed and consolidated this QPT~\cite{houck2012chip,toyoda2013experimental}.

As qubit-cavity coupling strength increases beyond weak coupling regime, the traditional Jaynes-Cummings description of qubit-cavity interaction becomes invalid, where the neglected counter-rotating terms (CRTs) dominates the physics~\cite{niemczyk2010circuit}.
Then, the Rabi-Hubbard model including CRTs,
is introduced to correctly describe the qubit-cavity coupling~\cite{zheng2011importance,schiro2012phase,schiro2013quantum,kumar2013quantum,flottat2016quantum}.
Based on the analytical and numerical exact methods, the phase diagram at ground state is significantly different from the counterpart in the JCH model~\cite{schiro2012phase,flottat2016quantum}.
It is mainly due to the fact that
continuous $U(1)$ symmetry in the JCH model is changed to discrete $Z_2$ symmetry in the Rabi-Hubbard model.
Hence, CRTs is considered crucial to reveal novel critical properties of the Rabi-Hubbard model~\cite{zheng2011importance}.
On the other hand, the collective effect has been analyzed based on the extended JCH model, termed as the RWA-Dicke-Hubbard model~\cite{lei2008quantum},
which describes interaction of multi-qubits with a single photonic mode at each site.
It was found that increasing qubits number in each cavity has intrigued novel many-body phenomena.
As is known, the Dicke model was originally applied to study collective spontaneous emission atom ensembles in quantum optics~\cite{dicke1954coherence},
resulting in the superradiant QPT~\cite{emary2003quantum,lambert2004entanglement}.
Hence, the question naturally arises that for the Dicke-Hubbard lattice, what is the influence of the interplay between the CRTs and collective effect on the behaviors of QPT and the corresponding phase diagram?

This paper aims to give a comprehensive picture of Dicke-Hubbard model including CRTs.
Based on the mean-field theory, we generalize the extended coherent state approach self-consistently to obtain the ground state behavior of delocalized order parameter and quantify the phase transition boundary.
The influence of CRTs on the phase diagram will be systematically analyzed, by comparing with the corresponding RWA case.
Moreover, the quantum fidelity is applied to detect the critical behavior of the Dicke-Hubbard model~\cite{gu2010fidelity}.
As known, quantum fidelity and the corresponding susceptibility have already been successfully to characterize QPT~\cite{gu2010fidelity,chen2008numerically,liu2009large}.
The paper is organized as follows:
In Sec. II, we describe the Dicke-Hubbard model including the CRTs.
In Sec. III, we develop a self-consistently extended coherent state approach within the mean-field framework.
The ground state energy and delocalized order parameter are obtained accordingly.
In Sec. IV, we discuss the localization-delocalization QPT, the corresponding phase boundaries, and the delocalization induced
concurrence enhancement.
Finally in Sec. V, we give a brief summary.

\section{Dicke-Hubbard lattice}
In this section, we firstly describe the Hamiltonian of the Dicke-Hubbard lattice.
Then, we generalize the extended coherent state approach within mean-field theory,
to self-consistently solve the mean-field Dicke-Hubbard model.

\subsection{Model}
The Dicke-Hubbard model, describing the interplay between intra-site qubit-cavity interaction and inter-site photon hopping, is given by~\cite{greentree2006quantum,hartmann2006strongly,hartmann2008quantum,lei2008quantum}
\begin{eqnarray}
\hat{H}=\sum_n\hat{H}^{\textrm{Dicke}}_n-{\kappa}\sum_{{\langle}n,m{\rangle}}\hat{a}^{\dag}_n\hat{a}_m.
\end{eqnarray}
The Hamiltonian $\hat{H}^{\textrm{Dicke}}_n$ shows collective interaction of $N$ identical two-level qubits with a single mode photonic field
in the $n$th cavity, expressed by the standard Dicke model~\cite{dicke1954coherence}
\begin{eqnarray}
\hat{H}^{\textrm{Dicke}}_n={\epsilon}\hat{J}^n_{z}+{\omega_0}\hat{a}^{\dag}_n\hat{a}_n+\frac{2\lambda}{\sqrt{N}}(\hat{a}^{\dag}_n+\hat{a}_n)\hat{J}^n_{x},
\end{eqnarray}
where $\hat{J}^n_{a}=\frac{1}{2}\sum_i\hat{\sigma}^{n,i}_{a}~(a=x,y,z)$ is the collective angular momentum operators with $\hat{\sigma}^{n,i}_a$ the Pauli operator of two-level qubit, and $\epsilon$ is the energy splitting.
$\hat{a}^{\dag}_n~(\hat{a}_n)$ creates (annihilates) one photon in the $n$th cavity, with the energy $\omega_0$.
$\lambda$ is the qubit-photon coupling strength, and $N$ is the qubits number within each cavity.
$\kappa$ is the inter-site photon hopping strength between nearest-neighbor sites $n$ and $m$, labeled by ${\langle}n,m{\rangle}$.
In this paper, we set $\hbar=1$ as unit for convenience.

For the finite $N$ Dicke system $\hat{H}^{\textrm{Dicke}}_n$ in the $n$th cavity, the parity operator $\hat{P}_n=\exp\{i\pi\hat{\Lambda}_n\}$ can be introduced, in terms of the total excitation number $\hat{\Lambda}_n=\hat{J}^n_z+\hat{a}^{\dag}_n\hat{a}_n$. Then, the Hilbert space of $\hat{H}^{\textrm{Dicke}}_n$ is separated into
two noninteracting subspace, due to the parity conservation as $[\hat{H}^{\textrm{Dicke}}_n,\hat{P}_n]=0$.
While for the Dicke-Hubbard model due to the photon hopping, parity $\hat{P}_n$ at local site $n$ appears broken.
Accordingly, the total system excitation number is given by
$\hat{\Lambda}=\sum_{n}(\hat{J}^n_z+\hat{a}^{\dag}_n\hat{a}_n)$, which results in the new parity operator $\hat{P}=\Pi_n\hat{P}_n$.
Thus, the orthogonal subspaces are properly re-arranged.
Moreover, the local Dicke model $\hat{H}^{\textrm{Dicke}}_n$ exhibits superradiant QPT in thermodynamic limit ($N\rightarrow\infty$),
at the critical qubit-cavity coupling strength $\lambda_c=\sqrt{\omega_0\epsilon}/2$, where the symmetry ($\hat{P}_n$) is broken~\cite{emary2003quantum,lambert2004entanglement}.
Below this critical strength ($\lambda<\lambda_c$), $\hat{H}^{\textrm{Dicke}}_n$ stays in the normal phase with microscopic excitation.
On the contrary, the local Dicke model shows superradiant phase as $\lambda>\lambda_c$, where the system particle number is macroscopically excited.

In the limiting case of $N=1$, where single qubit is embedded in each cavity, the Dicke-Hubbard model is reduced to the Rabi-Hubbard model~\cite{zheng2011importance,schiro2012phase,schiro2013quantum,kumar2013quantum,flottat2016quantum},
with the intra-site Hamiltonian modified to
$\hat{H}^{\textrm{Rabi}}_n={\epsilon}\hat{\sigma}^n_{z}/2+{\omega}\hat{a}^{\dag}_n\hat{a}_n/2+{\lambda}(\hat{a}^{\dag}_n+\hat{a}_n)\hat{\sigma}^n_{x}$.
Then, under RWA as traditionally considered, it is further changed to the JCH model~\cite{greentree2006quantum,hartmann2006strongly,hartmann2008quantum},
where the qubit-cavity interaction then becomes ${\lambda}(\hat{a}^{\dag}_n\hat{\sigma}^n_{-}+\hat{\sigma}^n_{+}\hat{a}_n)$.
It is widely known that Mott-insulating-to-superfluid phase transition and Mott-lobes are clearly observed in the JCH model,
due to the conservation of  polariton number~\cite{koch2009superfluid,schmidt2009strong,schmidt2010excitations,nietner2012ginzburg,you2014phase,bujnowski2014supersolid,hayward2015superfluid,hur2015many}.
And the JCH model is tightly related with the Bose-Hubbard model~\cite{fisher1989boson,bloch2008many,lv2015coulomb,hartmann2008quantum,koch2009superfluid,hur2015many}.
While the Rabi-Hubbard model exhibits quite different ground state features from the JCH model,
which is mainly due to CRTs of the qubit-cavity coupling  ${\lambda}(\hat{a}^{\dag}_n\hat{\sigma}^n_{+}+\hat{\sigma}^n_{-}\hat{a}_n)$~\cite{zheng2011importance,schiro2012phase}.
Therefore, it is desirable to exploit the interplay of CRTs and collective interaction in the Dicke-Hubbard lattice.
It should be noted that this fundamental light-matter interaction (CRTs) is reported to be able to play the role as effective chemical potential,
which can stabilize the finite density of correlated photons out of vacuum in equilibrium~\cite{schiro2012phase}.

\subsection{Extended coherent-state approach}
We firstly introduce the mean-field theory to simplify the Dicke-Hubbard lattice to an effective single-site case~\cite{hartmann2006strongly,koch2009superfluid}.
The order parameter is introduced as $\psi={\langle}\phi_G|\hat{a}|\phi_G{\rangle}$ with $|\phi_G{\rangle}$ the ground state,
to denote the QPT of photons in the Dicke-Hubbard model.
The Dicke-Hubbard system is in delocalized phase of photons for non-zero $\psi$ ($\psi{\neq}0$), and in localization phase for vanishing $\psi$
with a fixed number of particle excitations at each site.
Thus, the inter-site photon hopping term is decoupled as
$\hat{a}^{\dag}_n\hat{a}_m={\psi}^{*}\hat{a}_n+{\psi}\hat{a}^{\dag}_m-|\psi|^2$.
And the $n$th site Hamiltonian becomes
$\hat{H}_n=\hat{H}^{\textrm{Dicke}}_n-{z}\kappa({\psi}\hat{a}^{\dag}_n+{\psi}^{*}\hat{a}_n)+z{\kappa}|\psi|^2$,
where $z$ is the number of nearest-neighbor sites, e.g., $z=2$ for one-dimensional array.
In the present paper, we set $z=3$ for two-dimensional photonic lattice~\cite{hartmann2006strongly,lei2008quantum}, which actually does not affect our main results.
Moreover, the real order parameter is considered real $\psi=\psi^{*}$.
Therefore, the reduced model becomes site independent, resulting in the effective mean-field Hamiltonian
\begin{eqnarray}~\label{hmf}
\hat{H}_{\textrm{MF}}={\epsilon}\hat{J}_{z}+{\omega_0}\hat{a}^{\dag}\hat{a}+\frac{2\lambda}{\sqrt{N}}(\hat{a}^{\dag}+\hat{a})\hat{J}_{x}
-z\kappa{\psi}(\hat{a}^{\dag}+\hat{a})+z{\kappa}{\psi}^2.
\end{eqnarray}

The mean-field theory is known to be exact for infinite dimensional lattices, i.e., $z{\rightarrow}\infty$.
It has been extensively applied to study the JCH model and RWA-Dicke-Hubbard model, which shows consistent results with the numerical exact methods~\cite{rossini2007mott,pippan2009excitation}.
Moreover, the mean-field theory is also considered in the Rabi-Hubbard model,
in which the results agree with those within the functional and monte carlo approaches~\cite{schiro2012phase,flottat2016quantum},
even in strong $\kappa$ regime.
Hence, we believe the mean-field theory introduced in the present paper, is applicable to correctly exploit the QPT features of
the Dicke-Hubbard model.

For the finite-size Dicke model, it is known that only small qubits size can be numerically handled within conventional Fock state basis,
e.g., $N{\le}32$~\cite{lambert2004entanglement,emary2003quantum}.
While considering the extended coherent state approach, the ground state can be obtained accurately for large qubits size of the Dicke model~\cite{chen2008numerically,liu2009large}.
Thus, we generalize the extended coherent state approach in this paper, by including the order parameter $\psi$ in a self-consistent way.
Specifically, under the unitary transforming operator $\mathcal{\hat{R}}=e^{-i\frac{\pi}{2}\hat{J}_y}$,
the mean-field Hamiltonian $\hat{H}_{\textrm{MF}}$ at Eq.~(\ref{hmf}) is transformed to
\begin{eqnarray}
\hat{H}_R&=&{\mathcal{\hat{R}}^{\dag}}\hat{H}_{\textrm{MF}}{\mathcal{\hat{R}}}\\
&=&-\frac{\epsilon}{2}(\hat{J}_{+}+\hat{J}_{-})+{\omega_0}\hat{a}^{\dag}\hat{a}+
(\frac{2\lambda}{\sqrt{N}}\hat{J}_{z}-z\kappa{\psi})(\hat{a}^{\dag}+\hat{a})+z{\kappa}{\psi}^2,\nonumber
\end{eqnarray}
with $\hat{J}_{\pm}=\hat{J}_x{\pm}i\hat{J}_y$ is the angular momentum creating (annihilating) operator.
Accordingly, the wavefunction can be expanded in the Hilbert basis $\{|j,m{\rangle}{\otimes}|\phi{\rangle}_m\}$~($m=-j,-j+1,\cdots,j$)
with $\hat{J}_z|j,m{\rangle}=m|j,m{\rangle}$ and $j=N/2$,
where $|\phi{\rangle}_m$ is the photon part of the wavefunction corresponding to $|j,m{\rangle}$.
In sub-basis $|j,m{\rangle}$, the schr\"{o}dinger equation is specified by
\begin{eqnarray}~\label{ecsa}
&&-\frac{\epsilon}{2}(j^{+}_m|j,m+1{\rangle}+j^{-}_m|j,m-1{\rangle}){\otimes}|\phi{\rangle}_m\nonumber\\
&&+[{\omega_0}\hat{a}^{\dag}\hat{a}+(\frac{{2\lambda}m}{\sqrt{N}}-z{\kappa}\psi)(\hat{a}^{\dag}+\hat{a})+z{\kappa}\psi^2]|j,m{\rangle}{\otimes}|\phi{\rangle}_m
=E(\psi)|j,m{\rangle}{\otimes}|\phi{\rangle}_m,
\end{eqnarray}
where $j^{\pm}_m=\sqrt{j(j+1)-m(m{\pm}1)}$.
Through left-multiplying ${\langle}j,n|$ and introducing the displaced bosonic operator $\hat{A}_n=\hat{a}+g_n$ with $g_n=(\frac{{2\lambda}n}{\omega_0{\sqrt{N}}}-\frac{z{\kappa}{\psi}}{\omega_0})$,
we derive
\begin{eqnarray}
-\frac{\epsilon}{2}(j^{+}_n|\phi{\rangle}_{n+1}+j^{-}_n|\phi{\rangle}_{n-1})+{\omega_0}(\hat{A}^{\dag}_n\hat{A}_n-g^2_n+z{\kappa}\psi^2)
|\phi{\rangle}_n=E(\psi)|\phi{\rangle}_n.
\end{eqnarray}
It should be noted that the displaced coefficient $g_n$ depends on the undetermined order parameter $\psi$.
Then the bosonic wavefunction is expanded as
\begin{eqnarray}
|\phi{\rangle}_n=\sum^{Ntr}_{k=0}c_{n,k}|k{\rangle}_{{A}_{n}}=\sum^{Ntr}_{k=0}c_{n,k}\frac{(\hat{A}^{\dag}_n)^{k}}{\sqrt{k!}}|0{\rangle}_{{A}_{n}},
\end{eqnarray}
with $|0{\rangle}_{{A}_{n}}=e^{-g^2_n/2-g_na^{\dag}}|0{\rangle}_a$ and the bare ground state $a|0{\rangle}_a=0$.
Finally, we obtain $\psi$ dependent equation as
\begin{eqnarray}~\label{ee1}
&&{\omega_0}(l-g^2_n+z{\kappa}\psi^2)c_{n,l}
-\frac{\epsilon}{2}j^+_n\sum^{Ntr}_{k=0}c_{n+1,k}{_{A_{n}}}{\langle}l|k{\rangle}_{A_{n+1}} \nonumber\\
&&-\frac{\epsilon}{2}j^-_n\sum^{Ntr}_{k=0}c_{n-1,k}{_{A_{n}}}{\langle}l|k{\rangle}_{A_{n-1}}
=E({\psi})c_{n,l},
\end{eqnarray}
where ${_{A_{n+1}}}{\langle}l|k{\rangle}_{A_{n}}=(-1)^{k}D_{l,k}(g)$, with
$D_{l,k}(g)=e^{-g^2/2}\sum^{\min[l,k]}_{r=0}\frac{(-1)^r\sqrt{l!k!}g^{l+k-2r}}{(l-r)!(k-r)!r!}$
and $g=\frac{2\lambda}{\omega_0{\sqrt{N}}}$. In the present work, we set the photon truncation number $\textrm{Ntr}{\le}40$.
Then by self-consistently tuning the order parameter $\psi$, the ground state energy can be gained by $E_g=\min_{\psi}\{E(\psi)\}$ with relative error less than $10^{-5}$.
The corresponding order parameter $\psi$ is given by
\begin{eqnarray}~\label{psi1}
\psi&=&{\langle}\phi_G|\hat{a}|\phi_G{\rangle}
=\sum_{m,k,k^{\prime}}c^{*}_{m,k}c_{m,k^{\prime}}{_{A_{m}}}{\langle}k|\hat{a}|k^{\prime}{\rangle}_{A_{m}}\\
&=&\sum_{m,k,k^{\prime}}c^{*}_{m,k}c_{m,k^{\prime}}(\sqrt{k^{\prime}}\delta_{k,k^{\prime}-1}-g_m\delta_{k,k^{\prime}}).\nonumber
\end{eqnarray}
Specifically, we firstly find the global minimum of $E_{0}$, the corresponding ground state $|\phi_{0}{\rangle}$ and order parameter $\psi_{0}$ by numerically solving Eq.~(\ref{ee1}), in a wide range of input parameter $\psi$.
Next, we begin the iteration steps to self-consistently obtain the ground state energy and order parameter.
By including Eq.~(\ref{psi1}), the parameter $\psi_1={\langle}\phi_0|\hat{a}|\phi_0{\rangle}$ is obtained straightforwardly.
Then, by inserting $\psi_1$ into the Eq.~(\ref{ee1}), the lowest state $|\phi_1{\rangle}$ and the corresponding energy $E_1$ are gained.
Actually, $E_1$ is already very close to $E_0$.
We repeat such steps until the relative errors of both the ground state energy and order parameter within two successive iterations are less than $10^{-5}$.

\begin{figure}[tbp]
\begin{center}
\vspace{-2.2cm}
\includegraphics[scale=0.5]{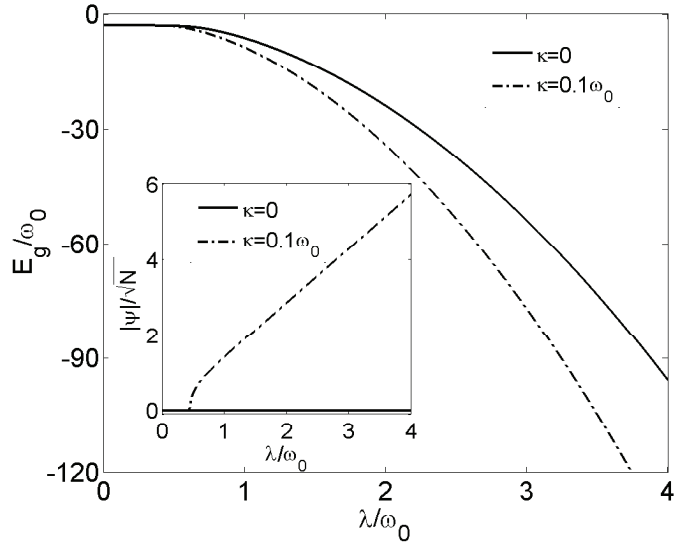}
\vspace{-3.2cm}
\end{center}
\caption{(Color online) Comparison of the ground state energy between the finite $N=6$ mean-field Dicke-Hubbard model at Eq.~(\ref{hmf}) ($\kappa=0.1\omega_0$) and the Dicke model in absence of inter-cavity hopping ($\kappa=0$).
The inset shows the rescaled order parameter $|\psi|/\sqrt{N}$ by tuning the qubits-cavity coupling strength.
The energy splitting is given by $\epsilon=\omega_0$.
}~\label{fig:fig0}
\end{figure}

Here, we apply the extended coherent state approach at Eq.~(\ref{ecsa}) to analyze the ground state energy of the finite qubits ($N=6$) Dicke-Hubbard model within the mean-field theory in Fig~\ref{fig:fig0} .
It is interesting to observe that in strong qubit-cavity coupling regime ($\lambda{\approx}0.5\omega_0$), the ground state energy including weak inter-site photon hopping ($\kappa=0.1\omega_0$)
becomes lower than the counterpart in absence of photon hopping process ($\kappa=0$).
It mainly results from the appearance of the delocalized phase exhibited by the inset of the Fig.~\ref{fig:fig0},
where the order parameter $\psi$ emerges accordingly.
This implies that the Dicke-Hubbard system prefers to stay at delocalized state of photons in strong qubit-cavity coupling regime.
While in absence of the inter-site photon hopping ($\kappa=0$), the mean-field Dicke-Hubbard Hamiltonian $\hat{H}_{\textrm{MF}}$ is reduced to the finite $N$ Dicke model, where $\psi$ always keeps zero due to the parity conservation~\cite{castanos2011no}.

\section{results and discussions}

\begin{figure}[tbp]
\begin{center}
\vspace{-2.2cm}
\includegraphics[scale=0.5]{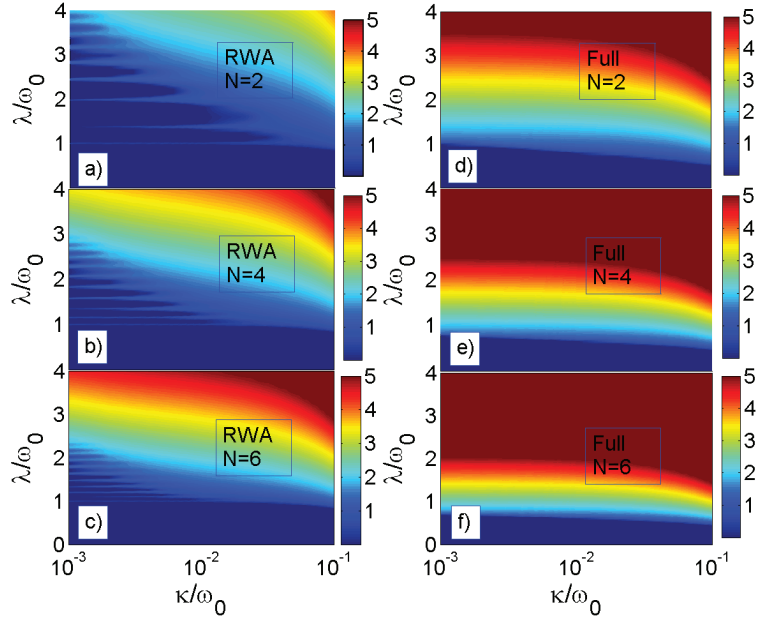}
\vspace{-3.2cm}
\end{center}
\caption{(Color online) Distribution of rescaled order parameter $|\psi|/\sqrt{N}$ of the  mean-field Dicke-Hubbard Hamiltonian at the ground state.
Various qubits sizes in each cavity are studied:
$N=2, 4, 6$ for a), b), c) within RWA, respectively;
$N=2, 4, 6$ for d), e), f) including CRTs, respectively.
The energy splitting is given by $\epsilon=\omega_0$.
}~\label{fig:fig1}
\end{figure}

\subsection{Localization-delocalization quantum phase transition}
We firstly investigate the simplest collective case of the Dicke-Hubbard model at Fig.~\ref{fig:fig1}, i.e., $N=2$.
It is known that within RWA, interaction of two qubits to the single photonic mode in the mean-field Hamiltonian at Eq.~(\ref{hmf}) is changed to
$\frac{\lambda}{\sqrt{2}}(\hat{a}^{\dag}\hat{J}_{-}+\hat{J}_{+}\hat{a})$.
Generally, it is difficult to analytically obtain the eigenspectrum of the mean-field Dicke-Hubbard model even under RWA, due to the existence of order parameter $\psi$.
However, if we focus on the Mott phase at resonance ($\epsilon=\omega_0$),
the eigen-solutions can be classified by the total excitation $\hat{\Lambda}=\hat{a}^{\dag}\hat{a}+\hat{J}_z+N/2$
under the Hilbert space $\{|n{\rangle}{\otimes}|j,m{\rangle}\}$.
The first class is the isolated single state $|0,0{\rangle}=|0{\rangle}{\otimes}|1,-1{\rangle}$ with energy $E_{0,0}=-\omega_0$.
The second class are single excited states $|\pm,0{\rangle}=(|1{\rangle}{\otimes}|1,-1{\rangle}{\pm}|0{\rangle}{\otimes}|1,0{\rangle})/\sqrt{2}$
with energy $E_{\pm,0}=\pm\lambda$.
For the excitation number $\Lambda{\ge}2$, the eigen-states are expressed as
\begin{eqnarray}~\label{efrwa}
|0,n{\rangle}&=&\frac{-\sqrt{n}|n+1{\rangle}{\otimes}|1,-1{\rangle}+\sqrt{n+1}|n-1{\rangle}{\otimes}|1,1{\rangle}}{\sqrt{2n+1}}\\
|\pm,n{\rangle}&=&\frac{\sqrt{n+1}|n+1{\rangle}{\otimes}|1,-1{\rangle}{\pm}\sqrt{2n+1}|n{\rangle}{\otimes}|1,0{\rangle}
+\sqrt{n}|n-1{\rangle}{\otimes}|1,1{\rangle}}{\sqrt{4n+2}},\nonumber
\end{eqnarray}
with the energy $E_{0,n}=\omega_0{n}$ and $E_{\pm,n}=\omega_0{n}{\pm}\lambda\sqrt{2n+1}~(n{\ge}1)$, respectively.
Therefore, the critical qubit-cavity coupling strength to quantify the superfluid-like phase transition can be obtained by
\begin{eqnarray}
\lambda^{0}_c&=&\omega_0\\
\lambda^{n}_c&=&\frac{\omega_0}{2}(\sqrt{2n+1}+\sqrt{2n-1})~n{\ge}1,\nonumber
\end{eqnarray}
due to the energy level crossings, which has been similarly obtained in Ref.~\cite{lei2008quantum}.
These transitions are clearly exhibited at Fig.~\ref{fig:fig1}(a), where Mott lobes separate the Mott-insulating-like phase and superfluid-like phase.
It should be noted that the analytical expressions of critical coupling strength are obtained based on the resonance condition ($\epsilon=\omega_0$),
resulting in $\lambda^n_c$ independent of $\kappa$ .
As more qubits are included, e.g., $N=4,~6$ at Figs.~\ref{fig:fig1}(b) and \ref{fig:fig1}(c), the number of Mott lobes also increases,
which implies frequent energy level crossings.
The superfluid order parameter is also strengthened~\cite{lei2008quantum}.
Hence, it can be concluded that increasing qubits number enriches behaviors of the phase diagram and QPT.

\begin{figure}[tbp]
\begin{center}
\includegraphics[scale=0.6]{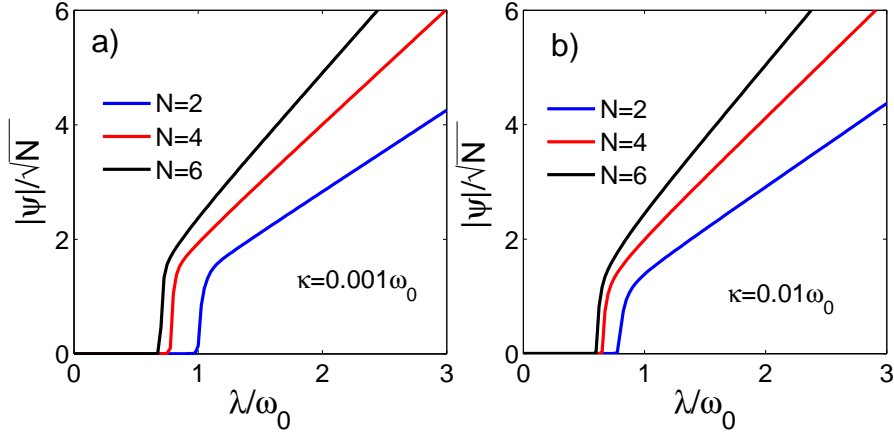}
\end{center}
\caption{(Color online) Behaviors of the rescaled delocalized order parameter $|\psi|/\sqrt{N}$ by tuning the qubit-cavity coupling strength, with various intra-cavity qubit numbers.
The energy splitting is given by $\epsilon=\omega_0$.
}~\label{fig:fig1-1}
\end{figure}

Then, we turn to study the influence of CRTs on $N=2$ mean-field Dicke-Hubbard model, shown at Fig.~\ref{fig:fig1}(d).
It is found that Mott lobes, exhibited under RWA at Fig.~\ref{fig:fig1}(a), are fully suppressed.
This result directly exhibits the crucial effect  of the CRTs on the elimination of Mott lobes, due to the change of system symmetry.
It originates from the fact that the $U(1)$ symmetry is broken.
Thus, the eigen-functions within RWA at Eq.~(\ref{efrwa}) are totally destroyed,
which are replaced by the semi-analytical solutions under the $Z_2$ symmetry~\cite{braak2011integrability,chen2012exact,he2015exact}.
It should be noted that such deviation is quite similar to the comparison made between the Rabi-Hubbard model and JCH model~\cite{schiro2012phase},
though the qubits number in each cavity is different.
Consequently, the signal of order parameter is strongly enhanced in strong qubit-cavity coupling regime ($\lambda{\gtrsim}\omega_0$),
mainly due to the CRTs exciting finite correlated-photons out of vacuum~\cite{schiro2012phase,schiro2013quantum,kumar2013quantum,flottat2016quantum}.
It is interesting to point out that by decreasing the inter-site hopping strength, the critical line separating localization and delocalized phases is stabilized, shown at Fig.~\ref{fig:fig1}(d).
This result is in sharp contrast to that in the Rabi-Hubbard model,
where the boundary line exhibits quick increase as the hopping strength decreases~\cite{schiro2012phase}.
We propose that this effect originates from the collective interaction of multi-qubits with the cavity photons.
It makes the critical behavior easier to observe compared to the Rabi-Hubbard case,
particularly in the weak inter-site hopping regime.
As the qubits number increases at Figs.~\ref{fig:fig1}(e) and \ref{fig:fig1}(f), the profiles of phase diagram are similar to the $N=2$ case.
Moreover, the signal of delocalization of photons shows monotonic enhancement by increasing intra-cavity qubit numbers, which is clearly demonstrated
at Figs.~\ref{fig:fig1-1}(a) and \ref{fig:fig1-1}(b).
Therefore, we believe these results under the influence of the CRTs are stable for collective-qubits Dicke-Hubbard model.

\subsection{Phase transition boundary}
\begin{figure}[tbp]
\begin{center}
\includegraphics[scale=0.5]{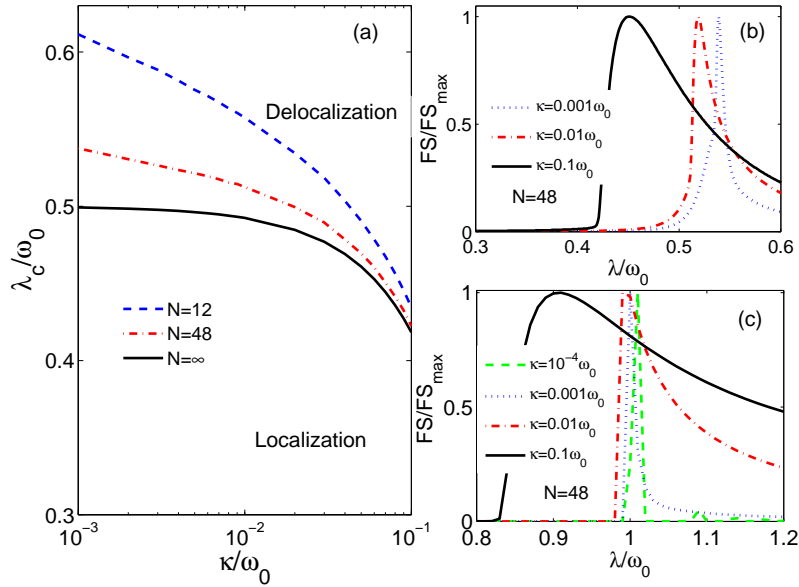}
\end{center}
\caption{(Color online)
(a) Mean-field phase transition boundary for the Dicke-Hubbard model with CRTs at finite qubits size ($N=12, 48$) and in thermodynamic limit ($N\rightarrow\infty$), respectively.
The critical qubit-cavity coupling strength $\lambda_c$ is exhibited as a function of inter-site photon hopping strength $\kappa$.
(b) The renormalized fidelity susceptibility $\textrm{FS}/\textrm{FS}_{\max}$ by including CRTs, specified at Eq.~(\ref{fs});
(c) The renormalized fidelity susceptibility with RWA.
The energy splitting is given by  $\epsilon=\omega_0$.
}~\label{fig:fig2}
\end{figure}

We firstly investigate the phase transition behavior at the ground state by including CRTs, in thermodynamic limit $N{\rightarrow}\infty$.
By applying the Holstein-Primakoff transformation, collective angular momentum operators are transformed to bosonic operators as
$\hat{J}_+=\hat{b}^{\dag}\sqrt{N-\hat{b}^{\dag}\hat{b}}$, $\hat{J}_-=\sqrt{N-\hat{b}^{\dag}\hat{b}}\hat{b}$,
and $\hat{J}_z=\hat{b}^{\dag}\hat{b}-N/2$, with the commutative relation $[\hat{b},\hat{b}^{\dag}]=1$.
It is known that in the superradiant phase, new bosonic operators with macroscopic displacements are introduced as
$\hat{c}=\hat{a}+\sqrt{N}\alpha$ and $\hat{d}=\hat{b}-\sqrt{N}\beta$,
whereas $\alpha=\beta=0$ in the normal phase~\cite{emary2003quantum}.
Through large $N$ expansion of the mean-field Hamiltonian $\hat{H}_{\textrm{MF}}$ with respect to bosonic operators $\hat{c}$ and $\hat{d}$,
the ground state energy can be obtained up to the order $O(N)$ as
\begin{eqnarray}~\label{eg}
\frac{E_G(\alpha,\beta)}{N}=(\omega_0-{z}\kappa)\alpha^2-4\lambda\alpha\beta\sqrt{1-\beta^2}+\epsilon(\beta^2-1/2).
\end{eqnarray}
By minimizing the ground state energy, we finally gain
\begin{eqnarray}
(\omega_0-z{\kappa})\alpha-2\lambda\beta\sqrt{1-\beta^2}&=&0,\\
2\alpha\lambda\sqrt{1-\beta^2}-\frac{2\alpha\lambda\beta^2}{\sqrt{1-\beta^2}}-\epsilon\beta&=&0,\nonumber
\end{eqnarray}
which results in
\begin{eqnarray}
\beta^2&=&\max\{0,\frac{1}{2}(1-\lambda^2_c/\lambda^2)\},\\
\alpha&=&\frac{2\lambda}{\omega_0}\beta\sqrt{1-\beta^2},\nonumber
\end{eqnarray}
with the critical transition point $\lambda_c=\sqrt{(\omega_0-z{\kappa})\epsilon}/2$.
It is clearly shown that the inter-site photon hopping reduces the critical qubit-cavity coupling strength, which identifies the normal-to-superradiant phase transition.
However, the appearance of delocalization ($\psi{\neq}0$) only requires the microscopic excitations.
Hence, the large $N$ expansion at Eq.~(\ref{eg}) to conserve the macroscopic excitations,
is only able to characterize order parameter in superradiant phase as $|\psi|=\alpha\sqrt{N}+O(1)$,
where the microscopic excitations can be safely ignored.

Then under the influence of CRTs, we turn to numerically investigate the critical qubit-cavity coupling strength for localization-delocalization phase transition in the large $N$ mean-field Dicke-Hubbard model, shown at Fig.~\ref{fig:fig2}(a).
It is surprising to observe that in the normal phase (${\langle}\hat{a}^{\dag}\hat{a}{\rangle}/N{\ll}1$),
the order parameter is negligible ($|\psi|{\ll}1$).
As the system goes into the superradiant phase with the photon field macroscopically excited ${\langle}\hat{a}^{\dag}\hat{a}{\rangle}/N{\sim}O(1)$,
the order parameter simultaneously rises into the finite value as $|\psi|{\sim}O(\sqrt{N})$, which is consistent with the result in thermodynamic limit.
To confirm such phase transition correspondence, we apply the renormalized fidelity susceptibility $\textrm{FS}/\textrm{FS}_{\max}$ to detect this criticality~\cite{gu2010fidelity},
where the fidelity susceptibility is given by
\begin{eqnarray}~\label{fs}
\textrm{FS}=2[1-F(\lambda)]/(\delta\lambda)^2,
\end{eqnarray}
with $\textrm{FS}_{\max}=\max\{\textrm{FS}\}$ and the fidelity $F(\lambda)=|{\langle}\phi_G(\lambda)|\phi_G(\lambda+\delta\lambda){\rangle}|^2$~ ($\delta\lambda{\ll}\lambda$).
It is found that for various inter-site photon hopping strengths, single peak exists at the critical point, shown at Fig.~\ref{fig:fig2}(b).
Hence, we propose that the boundary of normal-superradiant phase transition may overlap with the counterpart for localization-delocalization phase transition, within the framework of mean-field theory.
Moreover, as a comparison, we show behaviors of the renormalized fidelity susceptibility of Dicke-Hubbard model within RWA at Fig.~\ref{fig:fig2}(c).
The single peak also appears with the same photon hopping strengths as those at Fig.~\ref{fig:fig2}(b).
However, if photon hopping strength becomes further smaller, e.g., $\kappa=10^{-4}\omega_0$, multi-peaks emerges to demonstrate existence of Mott lobes,
as similarly shown at Figs.~\ref{fig:fig1}(a-c).
This implies that increasing intra-cavity qubit numbers will only shrink localization regime, but unable to fully suppress Mott lobes.

\section{Conclusion}

To summarize, we have studied quantum phase transition in the Dicke-Hubbard model within the mean-field theory, where CRTs
are included~\cite{schiro2012phase,schiro2013quantum}.
The extended coherent state approach has been generalized to numerically solve the ground state phase diagram with respect to self-consistency.
In the mean-field based RWA-Dick-Hubard model, the Mott lobes are clearly exhibited,
and critical qubit-cavity coupling strength to characterize Mott-insulating-to-superfluid-like phase transition has been analytically obtained.
While under the influence of the CRTs, the Mott lobes are fully suppressed,
mainly due to the fact that the system symmetry is re-established from $U(1)$ type
in the localized phase ($\psi=0$) to $Z_2$ type in the delocalized phase ($\psi{\neq}0$).
This new parity symmetry ($Z_2$) will stabilize finite density of correlated photons even at ground state.
Accordingly, phase transition pattern is significantly changed.
Moreover, the critical coupling strength $\lambda_c$ to separate localized and delocalized phases approaches a stable value by decreasing inter-site photon hopping strength.
 It is in sharp contrast to the counterpart in the Rabi-Hubbard model, where $\lambda_c$ increases dramatically by weaken inter-site photon hopping strength~\cite{schiro2012phase,schiro2013quantum}.
Hence, we conclude that this effect is mainly from the indirect qubit-qubit collective interaction mediated by single mode photonic field.
Then, we quantified the localization-delocalization transition boundary in finite $N$ mean-field Dicke-Hubbard model.
It is proposed that within mean-field framework, normal-superradiant phase transition boundary may overlap with that for localization-delocalization transition.
This result is further confirmed by the fidelity susceptibility, where single global peak is exhibited at critical coupling strength.
Hence, we believe that quantum fidelity may be utilized to measure quantum phase transition of the Dicke-Hubbard lattice.

Finally, we note that the dissipation effect has recently been analyzed in JCH model, which suppresses the delocalized regime~\cite{you2014phase,de2015effects}.
The extended coherent state approach can be applied and generalized to
exploit the novel behavior from the interplay between loss mechanism and CTRs in the Dicke-Hubbard model,
which we will pursue in the future work.

\section{acknowledgement}
We Thank Mr. Shu He and Prof. Qing-Hu Chen  for helpful discussions.
This work was supported by the  National Natural Science Foundation of Special Theoretical Physics under Grant No. 11547124.
Chen Wang has been partially supported by the National Natural Science Foundation of China under Grant Nos.
11574052 and 11504074.



\end{document}